\documentclass{aa}
\bibpunct{(}{)}{;}{a}{}{,} % to follow the A&A style
\usepackage[utf8]{inputenc}

\usepackage{graphicx}
\usepackage[varg]{txfonts}

\usepackage{multicol}

\begin{document}

\title{Radio observations of massive stars in the Galactic centre: The Arches Cluster}
\author{Gallego-Calvente\inst{1}\fnmsep\thanks{\email{gallego@iaa.es}}, A.~T.
        \and
         Sch\"odel\inst{1}, R.
        \and
         Alberdi \inst{1}, A.
        \and
         Herrero-Illana\inst{2}, R.
        \and 
         Najarro\inst{3}, F.
        \and
         Yusef-Zadeh \inst{4}, F.
        \and
         Dong, H.
        \and
         Sanchez-Bermudez\inst{5, 6}, J.
        \and
         Shahzamanian\inst{1}, B.
        \and
         Nogueras-Lara\inst{6}, F.
        \and 
         Gallego-Cano\inst{7}, E. 
        }
         
\institute{Instituto de Astrof\'isica de Andaluc\'ia (IAA-CSIC), Glorieta de la Astronom\'ia s/n,
              18008 Granada, Spain e-mail: gallego@iaa.es
         \and
             European Southern Observatory (ESO), Alonso de C\'ordova 3107, Vitacura, Casilla 19001, Santiago de Chile, Chile
         \and 
             Centro de Astrobiolog\'ia (CSIC/INTA), Ctra. de Ajalvir Km. 4, 28850 Torrej\'on de Ardoz, Madrid, Spain
         \and 
                 CIERA, Department of Physics and Astronomy Northwestern University, Evanston, IL 60208, USA
             \and 
                 Instituto de Astronom\'ia, Universidad Nacional Aut\'onoma de M\'exico, Apdo. Postal 70264, Ciudad de M\'exico 04510, M\'exico
         \and 
                 Max-Planck-Institut f\"ur Astronomie, K\"onigstuhl 17, Heidelberg, D-69 117, Germany
         \and 
            Centro Astron\'omico Hispano-Alem\'an (CSIC-MPG), Observatorio Astron\'omico de Calar Alto, Sierra de los Filabres, 04550, G\'ergal, Almer\'ia, Spain
             }
\date{}

\abstract{
    % methods
    We present high-angular-resolution radio observations of the Arches cluster in the Galactic centre, one of the most massive young clusters in the Milky Way. The data were acquired in two epochs and at 6 and 10\,GHz with the Karl G. Jansky Very Large Array (JVLA). The rms noise reached is three to four times better than during previous observations and we have almost doubled the number of known radio stars in the cluster. Nine of them have spectral indices consistent with thermal emission from ionised stellar winds, one is a confirmed colliding wind binary (CWB), and two sources are ambiguous cases. Regarding variability, the radio emission appears to be stable on timescales of a few to ten years. Finally, we show that the number of radio stars can be used as a tool for constraining the age and/or mass of a cluster and also its mass function.
}


\maketitle

\section{Introduction}

Massive stars are of fundamental importance to understanding star formation and galaxy evolution because of their key role in stirring and enriching the interstellar medium through intense ionising radiation, stellar winds, and supernovae. A striking feature of massive stars is their high fraction of multiplicity ($\sim$\,91\,$\%$, \citealt{sana:2015aa}), which must therefore be taken into account when studying their evolution and the heavy end of the initial mass function. Considering that the wind properties of massive stars are still poorly observationally constrained \citep[see][]{benaglia:2010aa}, radio observations can provide further information about the wind properties and multiplicity of the most massive stars.

As concerns the properties of radio stars, for isolated stars we expect a spectral index of $\alpha\approx0.6$ (flux density $S_{\nu}\propto\nu^{\alpha}$) arising from the thermal emission of the ionised wind, while binaries can show flat to inverted spectra ($\alpha\lesssim0$) which result from the contribution of non-thermal emission in colliding wind regions, provided this emission is not absorbed by the surrounding ionised wind, which is optically thick \citep{benaglia:2010aa,montes:2015aa}. Therefore, identifying binaries through radio observations is typically limited to the detection of wide binaries with periods of one year or longer \citep{sanchez-bermudez:2019aa}. In the case of short-period ($P\,\sim<\,1$\,yr) colliding wind binaries (CWBs), the stars are too close and the wind--wind collision region (WCR) is likely to lie within the optically thick region of the winds. Therefore, in these systems, only the free-free thermal emission from the unshocked winds is thought to be detected, thereby masking any effect of their binarity. However, for some short-period systems, the non-thermal emission escapes the absorption, contributing to a composite spectrum with a flat spectral index at certain orbital phases in systems with significant eccentricity. On the other hand, theoretical studies suggest that the free-free thermal emission from the WCR may also affect the total radio spectrum, a phenomenon that becomes more important as the stars get very close, increasing the spectral index to values steeper than the nominal ionised wind at millimeter wavelengths \citep{montes:2015aa}.

Past radio observations of massive stars lacked the sensitivity to reach beyond at most a few kiloparsecs. Because of the rareness and large mean distances of massive stars and clusters, researchers could were therefore only able to study a limited sample of such targets \citep{lang:2001aa, lang:2005aa, benaglia:2010aa, yusef-zadeh:2015aa}.

The Galactic centre (GC) region occupies less than 1\% of the volume of the Milky Way disc, but emits on the order 10\% of its total Lyman continuum flux, which is produced by a high number of massive, hot stars \citep[e.g.][]{figer:2004aa,mauerhan:2010aa,dong:2011aa}. The Arches, Quintuplet, and the Central Parsec massive young clusters lie within 30\,pc in projection of the central black hole Sagittarius\,A* (Sgr\,A*) and each contain $\gtrsim10^{4}$\,M$_{\odot}$ of stars that formed 2-6\,Myr ago \citep{figer:2004aa}. Massive young stellar objects (YSOs) and H\,II regions throughout the GC are further witnesses to recent or currently ongoing massive star formation (SF) throughout the GC \citep[e.g.][]{yusef-zadeh:2009aa,mauerhan:2010aa,nandakumar:2018aa,shahzamanian:2019aa}. In summary, the GC is the Milky Way's most important star forming region. The conditions in the GC resemble those in high-redshift starburst galaxies \citep{kruijssen:2013aa}. Thus, the GC is of special importance for studies of massive star formation.

However, there are some unique observational challenges for studying the stellar population of the GC. Interstellar extinction is extremely high ($A_{V}>30$\,mag) and also variable on arcsecond scales \citep[e.g.][]{scoville:2003aa,schoedel:2010aa,fritz:2011aa,nogueras-lara:2018aa,nogueras-lara:2019aa,nogueras-lara:2020aa}. Therefore, even in the near-infrared ($A_{K}\approx2.5$\,mag), stellar colours are dominated by reddening. Finally, sub-arcsecond angular resolution (at least as good as about 0.2\arcsec) is needed to overcome the high source crowding and to reliably study individual stars, which requires the use of the Hubble Space Telescope, or speckle or adaptive optics techniques from the ground.

% FIGURE 1
\begin{figure*}[th!]
    \centering
    \includegraphics[width=.9\textwidth]{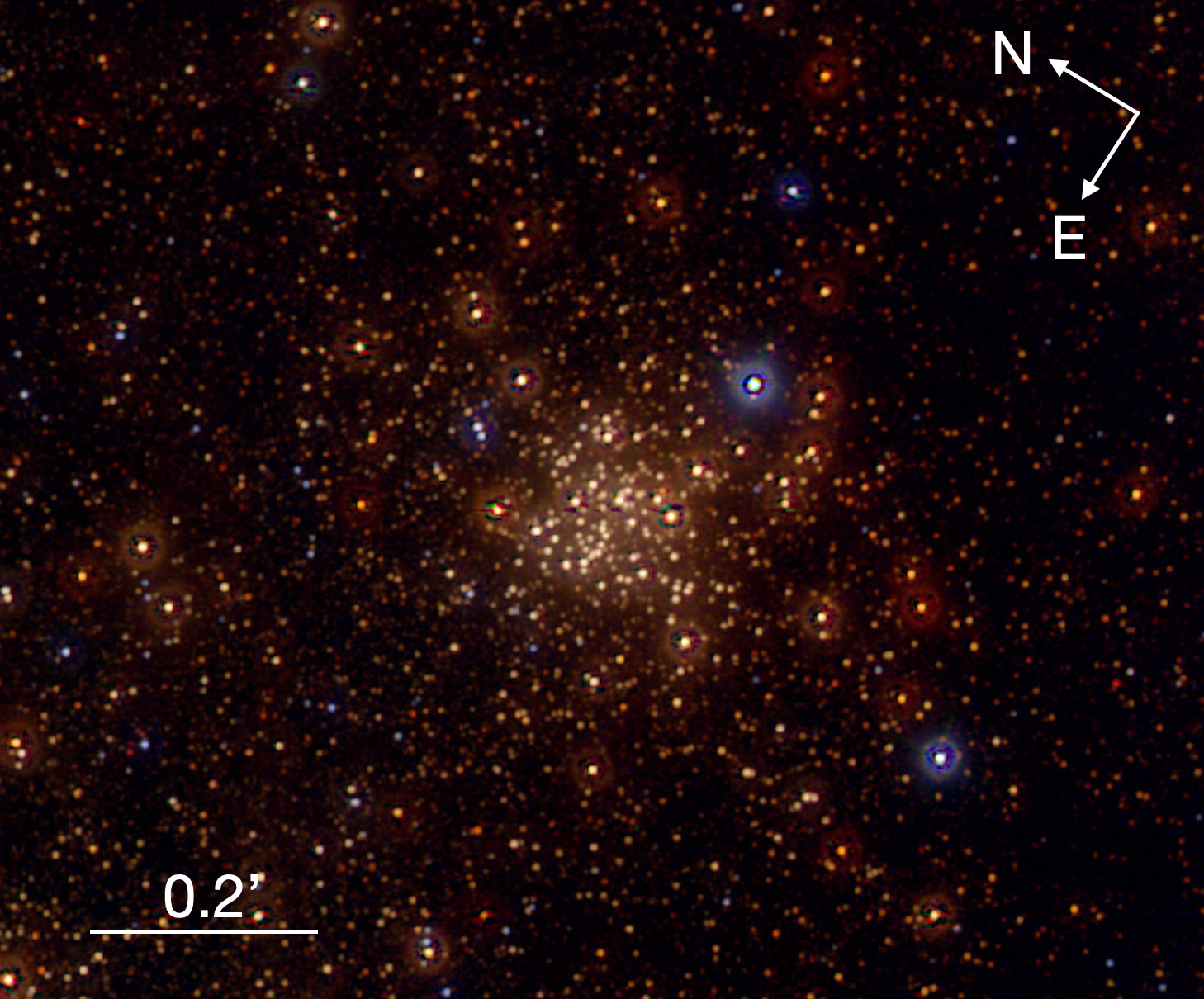}
    \caption{JHKs false colour image of the Arches cluster from the GALACTICNUCLEUS survey \citep{nogueras-lara:2018aa, nogueras-lara:2019ab}.}
    \label{fig:arches}
\end{figure*}

In this work, we focus on the Arches cluster that is located at a projected distance of $\sim$\,26\,pc to the northeast of Sgr\,A* (11\arcmin, angular distance). It contains a few $\times\,10^{4}$\,M$_{\odot}$ with more than 100 O-stars and thus belongs to a small handful of young, massive starburst-like clusters known in the Milky Way (in addition to, e.g. NGC3603, Quintuplet, or Westerlund 1 and 2). The cluster was formed between 2 and 4\,Myr ago \citep[e.g.][]{figer:1999aa,figer:1999ab,najarro:2004aa,clarkson:2012aa,clark:2018aa,clark:2019aa} and has a half-light radius of $r_{\textrm{h}}\approx0.48$\,pc \citep{hosek:2015aa}.

\citet{clark:2018aa, clark:2019aa} recently provided the spectral classifications for 88 Arches cluster members based on a new multi-epoch near-infrared (NIR) spectroscopic survey of the cluster obtained with the integral field spectrograph SINFONI mounted on the Very Large Telescope (VLT) and photometry from the Hubble Space Telescope Wide Field Camera 3 (HST/WFC3), increasing the number of cluster members found by previous studies from \citet{martins:2008aa} by a factor of about three.

\citet{lang:2005aa} carried out a multi-frequency, multi-configuration, and multi-epoch study of the Arches cluster with the Very Large Array (VLA), reaching $1\,\sigma$ flux uncertainties of $\sim$\,30\,$\mu$Jy at 8.5\,GHz. They detected ten compact sources in the Arches cluster that could all be associated with massive, young stars detected in the NIR.
All but the brightest of these sources (AR1) were unresolved and their spectral indices were consistent with stellar wind sources, which were used to estimate their mass-loss rates. Roughly half of the sources showed indications of moderate variability in their work.

In the present study, we revisited the Arches cluster aiming to pick up the thermal and non-thermal emission from the ionised gas in the outer wind regions of young, massive stars. We took advantage of the significantly increased sensitivity of the Karl G. Jansky Very Large Array (JVLA), thanks to the increase of the bandwidth and the implementation of a new correlator since the previous observations.
Additionally, recent spectroscopic work in the NIR \citep{clark:2018aa,clark:2019aa} provided us with reliable constraints on the properties of the stars.

% End of Introduction section

\section{Observations and imaging}

% TABLE 1
\begin{table}[!t]
  \centering
        \caption{Observational properties}
        \footnotesize
    \setlength{\tabcolsep}{4pt}
        \begin{tabular}{c c c c}
                Observation & Band\tablefootmark{a} & JVLA          & On source time     \\
                date        &                       & configuration & (minutes)          \\
                \noalign{\smallskip}
                \hline \hline
                \noalign{\smallskip}
                   Oct 04, 2016  & X & A & 55 \\
               Oct 26, 2016  & X & A & 55 \\
               Apr 11, 2018  & X & A & 55 \\
               Jun 10, 2018  & C & A & 74 \\
        \end{tabular}
        \tablefoot{
        \tablefoottext{a}{Frequency range of 4-8\,GHz for C-band and 8-12\,GHz for X-band. Therefore, the total bandwidth was 4\,GHz on each band. The number of spectral windows was 32 and the number of channels 64 in both cases.}
        }
        \label{table:obs}
\end{table}

% TABLE 2
\begin{table*}[!t]
        \caption{Properties of the images}
        \footnotesize
    \setlength{\tabcolsep}{5.0pt} % Default value: 6pt
        \begin{tabular}{c c c c r c c c}
        \centering
          Epoch & Band & Frequency \tablefootmark{a} & Synthesised beam & \multicolumn{1}{r}{P. A.\tablefootmark{b}}  & rms noise\tablefootmark{c}  & $(u,v)$ cut-off  & Largest angular scale\tablefootmark{d}  \\
                &      & (GHz)                       & (arcsec $\times$ arcsec)             & \multicolumn{1}{c}{(degrees)}               & ($\mu$Jy\,beam$^{-1}$)              & (k$\lambda$)    & (arcsec)                                \\
          \noalign{\smallskip}
          \hline \hline
          \noalign{\smallskip}
          2016 & X & 10.0  &  $0.48 \times 0.18$  &    20.84   &  2.5  & 150  & 5.3 \\
          2018 & X & 10.0  &  $0.46 \times 0.17$  & $-$17.33   &  4.0  & 150  & 5.3 \\
          2018 & C &  6.0  &  $0.62 \times 0.28$  &     2.82   &  4.7  & 100  & 8.9 \\
        \end{tabular}
        \tablefoot{\smallskip \\
         \tablefoottext{a}{Representative frequency, in Gigahertz.}\\
         \tablefoottext{b}{The position angle (P. A.) of the fitted major axis for the synthesised beam, in degrees.}\\
         \tablefoottext{c}{Off-source root mean square noise level reached.} \\
         \tablefoottext{d}{Scale at which severe attenuation of large-scale structure occurs (in arcseconds).} 
         }
         \label{table:prop}
\end{table*}

We observed the radio continuum emission from the Arches cluster using the JVLA of the National Radio Astronomy Observatory (NRAO)\footnote{The NRAO is a facility of the National Science Foundation (NSF) operated under cooperative agreement by Associated Universities, Inc.}. The position $\alpha$, $\delta_{\textrm{(J2000)}}$ = 17$^{\textrm{h}}$\,45$^{\textrm{m}}$ \,50.49$^{\textrm{s}}$, $-$28\degr\,49\arcmin\,19.92\arcsec\, was taken as phase centre. There are three epochs of X-band (3.0\,cm or a representative frequency of 10\,GHz) and one of C-band (5.0\,cm or 6\,GHz) observations. Details are listed in Table\,\ref{table:obs}.
All the observations in all epochs were taken in the A configuration to achieve the highest angular resolution. This configuration also helped us to filter out part of the extended emission from the Arched Filaments (G0.10+0.08), a very extended H\,II region in which the Arches cluster is embedded.

At all frequencies, J1744$-$3116 was used as a phase calibrator and J1331+305 (3C286) as a band-pass and flux density calibrator.
For C- and X-bands, we required a sensitivity of 3.0\,$\mu$Jy\,beam$^{-1}$.
Raw data were processed automatically through the JVLA calibration pipeline performing an initial flagging and calibration. Extra flagging was necessary to remove the lost or corrupted data.
We performed standard data reduction using the Common Astronomy Software Applications package (CASA) developed by an international consortium of scientists\footnote{Scientists based at the National Radio Astronomical Observatory (NRAO), the European Southern Observatory (ESO), the National Astronomical Observatory of Japan (NAOJ), the Academia Sinica Institute of Astronomy and Astrophysics (ASIAA), the CSIRO division for Astronomy and Space Science (CASS), and the Netherlands Institute for Radio Astronomy (ASTRON) under the guidance of NRAO.}.

Images were created with CASA using the classical task \textsc{clean} in interactive mode. We also tested a more advanced form of imaging, \textsc{multi-scale clean}, to distinguish between the point sources and the extended emission. \textsc{multi-scale clean} involves the use of multiple scales by means of an extension of the classical \textsc{clean} algorithm assuming the sky is composed of emission at different angular scales. Its use did not improve the quality of the image. We also probed various other options for the task \textsc{clean} including wide-field, multi-term, multi-frequency synthesis and with w-projection without resulting in any significant improvements. Additionally, we checked different weighting schemes to correct for visibility sampling effects. Natural weighting produced an optimum image with the lowest noise (highest signal-to-noise ratio (S/N)). The gain parameter, which sets the fraction of the flux density in the residual image that is removed and placed into the clean model at each minor cycle iteration, was set to 0.05 to help when cleaning our image with diffuse emission.
Visibility data sets from 2016 were cleaned in two ways, separately and concatenated. In the first case, self-calibration could not be applied, possibly because this method requires sufficient S/N at each solution interval. In the second case (2016 data sets concatenated), self-calibration was applied just in phase, just in amplitude, and in phase and amplitude simultaneously. The best image was a result of applying one cycle of phase self-calibration to correct for antenna-based phase errors selecting interactively the brightest compact sources as input model, as well as amplitude self-calibration (the latter with no changes in the flux density of the compact sources). The off-source root mean square (rms) noise level reached in the final image with this procedure was 2.5\,$\mu$Jy\,beam$^{-1}$. 
Self-calibration could not be applied to the 2018 single data sets either. The off-source thermal noise reached in this epoch was $4.0$ and 4.7\,$\mu$Jy\,beam$^{-1}$ for X- and C-bands, respectively.
All the final images were primary beam corrected to account for the change in sensitivity across the primary beam.
Table\,\ref{table:prop} summarises the properties of the final images.

As the cluster is located in an extended H\,II region and the primary beam sizes at 5 and 3\,cm were relatively large (approximately 5\arcmin and 8\arcmin, respectively), a spatial frequency cut-off was also required in order to resolve the diffuse emission and detect the compact sources with high S/N. Considering that the field of view (FoV) is of the order of $\theta\textrm{(PB)} = 42\,\nu_{\textrm{GHz}}^{-1}$ in arcminutes (i.e. 8.4 and 4.2 arcminutes at C- and X- bands respectively), the Central Parsec Cluster and Sgr\,A* are not within the FoV at any frequency. Thus, no flux from Sgr\,A* is bleeding into the target fields. 
Nevertheless, we verified via large FoV imaging that Quintuplet was within the FoV for both epochs but does not contribute significantly to the flux density. Therefore, we are certain that no flux from nearby bright regions leaked into the FoV of our reconstructed images. 
Table\,\ref{table:prop} also lists the cut-offs made to the data in the $(u,v)$ plane and the scale at which severe attenuation of large-scale structure occurs.

% End of Observations and Imaging section

\section{Results}

\subsection{Point source detection and calibration}

The positions and flux densities of the detected radio sources were taken from the final primary-beam-corrected images using the \textsc{CASA-imfit} task which fits elliptical 2D Gaussians to interactively selected polygon regions around source candidates. The task provided us with estimations of the positions of the maxima, the total flux densities, and the errors of these values taking the quality of the fit and the image rms into account. Uncertainties in positions were calculated by adding in quadrature a systematic error of 0.05\arcsec \citep{dzib:2017aa} to the formal error of the fit, 0.5\,$\theta$/SNR \citep{reid:1988aa}, where $\theta$ is the beam size and SNR the S/N. The systematic error accounts for the thermal noise and uncertainties introduced by the phase calibration process. In the determination of the flux-density uncertainties, we also considered the percentage in the calibration error of the peak flux densities at the frequencies observed \citep{perley:2013aa} and a factor that takes into account whether a source is resolved or unresolved. With these premises, 18 and 15 sources were detected above five times the off-source rms noise level at 10\,GHz and 6\,GHz, respectively.

% FIGURE 2
\begin{figure*}[th!]
    \centering
    \includegraphics[width=.9\textwidth]{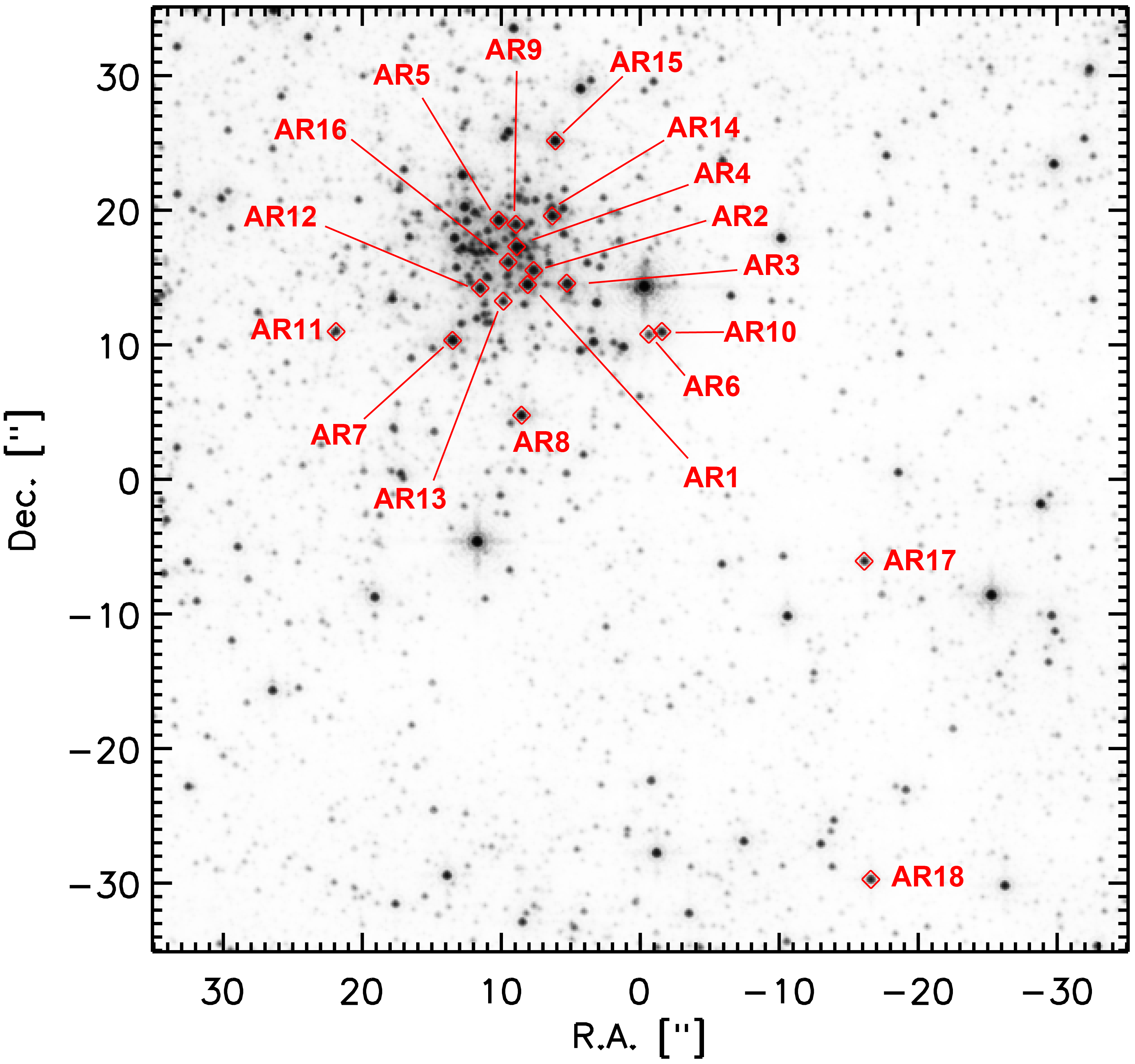}
    \caption{HST/WFC3 F153M image of the Arches cluster with identified radio stars labelled.}
    \label{fig:sources}
\end{figure*}

The 2016 X-band image has the highest S/N of all images and provides the most complete list of point sources. The presence of ionised gas in clouds of variable compactness, as well as artefacts introduced by the brightest radio sources into the images, can give rise to point sources that are either spurious or related to features within the extended emission. We therefore compared the positions of the radio point sources detected in the 2016 X-band image with the positions of stars in an HST/WFC3 F153M image of the Arches cluster \citep{hosek:2015aa} and eliminated all radio sources that did not show unambiguous coincidence (within less than half of the NIR and radio beam FWHMs). The radio stars are labelled in the NIR image of the Arches cluster shown in Fig.\,\ref{fig:sources}. Point sources were selected interactively in the 2016 radio image, restricting ourselves to those at $5\,\sigma$ above the off-source rms noise level. Within the approximately $2.1\arcmin\times 2.3\arcmin$ WFC3 image there are 18 of a total of 23 $5\,\sigma$ radio detections that clearly coincide with stars. With the 2018 data, we only detected the radio stars with the $5\,\sigma$ criterion (almost all the stellar point sources identified in 2016) and no spurious sources.

% FIGURE 3
\begin{figure*}[!t]
    \centering
    \includegraphics[width=.645\textwidth]{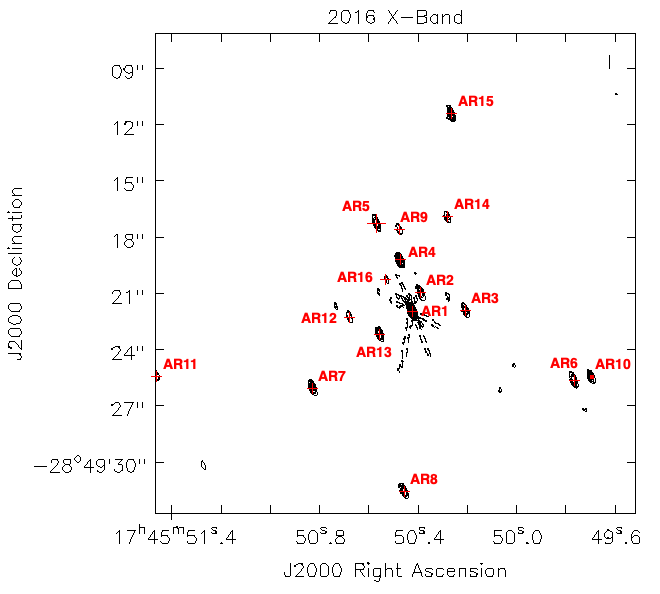}
    \includegraphics[width=.645\textwidth]{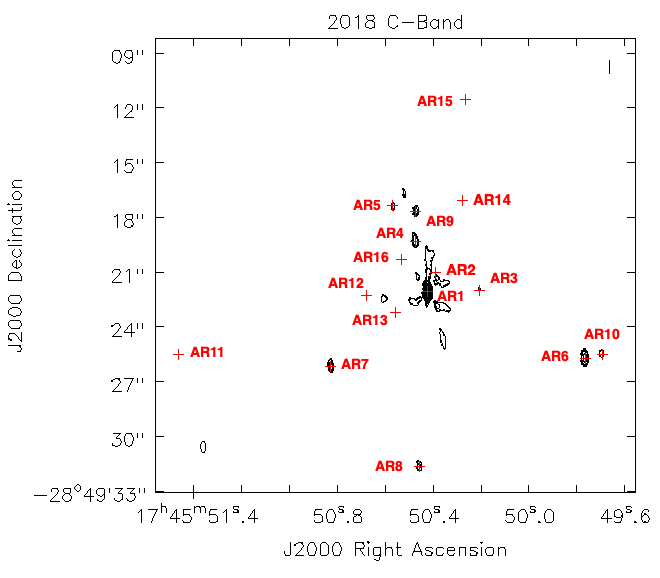}
    \caption{Top: Closeup onto the Arches cluster from the 2016 X-band image not corrected for primary beam attenuation. The clean beam is $0.48\arcsec \times 0.18\arcsec$, P.A. = $20.84\degr$. The off-source rms noise level is 2.5 $\mu$Jy beam$^{-1}$. The contour levels represent -1, 1, 3, 5, 10, 15, 25, 35, 45, 55, and 60 times the off-source rms noise level multiplied by 5.
    Bottom: Closeup onto the Arches cluster showing most of the detected sources from the 2018 C-band image not corrected for primary beam attenuation. The resolution is $0.62\arcsec \times 0.28\arcsec$, P.A. = $2.82\degr$. The off-source rms noise level is 4.7 $\mu$Jy beam$^{-1}$. The contour levels represent 3, 5, 10, 15, 25, 35, 45, 55 and 60 times the off-source rms noise level multiplied by 3.}
    \label{fig:Arches_X}
\end{figure*}

Figure\,\ref{fig:Arches_X} shows a closeup onto the 2016 X-band image and a closeup onto the 2018 C-band image of the Arches cluster with all radio stars labelled. Sources AR1-10 were reported by \citet{lang:2005aa}. Sources with higher numbers are our new detections.

When comparing the flux densities from our 2016 and 2018 X-band images, we noted an offset of a factor of about two between the epochs. As the observations by \citet{lang:2005aa} were done at $8.5$\,GHz, very close in frequency to our $10$\,GHz observations, we can compare our measurements with theirs to infer the systematic errors. Table\,\ref{table:point_sources} lists the measured flux densities for the point sources in the X-band image of the 2018 epoch along with the 8.5\,GHz fluxes of \citet{lang:2005aa}, where available. We should mention that this factor of about two between our 2016 and 2018 epochs is not due to a problem in the absolute flux density calibration of the observations, because the flux density for the amplitude calibrator is the same for both epochs. We can also discard that it is related with the self-calibration procedure, because it does not change the flux density of the compact sources. We performed several tests, obtaining images with and without self-calibration, and concatenating or treating the 2016 data sets separately, but the results were the same. We also used different options for the \textsc{clean} algorithm, including wide-field clean, or multi-term multi-frequency synthesis with identical finding. We could not identify the cause of this systematic error in our data reduction, but performed a secondary calibration as follows.

% TABLE 3
\begin{table*}[!th]
  \centering
        \caption{Detected point sources in the two X-band epochs and their flux density comparison to point sources at $8.5$\,GHz reported by \citet{lang:2005aa}.}
%       \footnotesize
        \setlength{\tabcolsep}{9pt} % Default value: 6pt
        \begin{tabular}{llllcc}
         \hline
         \noalign{\smallskip}
         Source & \multicolumn{1}{c}{R.A.\tablefootmark{a}}    & \multicolumn{1}{c}{Dec.\tablefootmark{a}} & \multicolumn{1}{c}{$S_{\textrm{8.5\,GHz}}$}  &  $S_{\textrm{X, 2018}}$  &  $S_{\textrm{X, 2016, re-calibrated}}$ \\
     \noalign{\smallskip}
                & \multicolumn{1}{c}{(J2000.0)}                & \multicolumn{1}{c}{(J2000.0)}             & \multicolumn{1}{c}{(mJy)}                    & (mJy)                    & (mJy)  \\
     \noalign{\smallskip}        
     \hline \hline
         \noalign{\smallskip}
        AR1   &  17 45 50.42 $\pm$ 0.05  &  $-$28 49 21.95 $\pm$ 0.05  & $1.90 \pm 0.05$\tablefootmark{b}    & $2.1  \pm 0.3$     &  $2.2  \pm 0.5$   \\
        AR2   &  17 45 50.39 $\pm$ 0.05  &  $-$28 49 20.93 $\pm$ 0.05  & $0.25 \pm 0.03$                     & $0.24 \pm 0.04$    &  $0.23 \pm 0.05$  \\ 
        AR3   &  17 45 50.21 $\pm$ 0.05  &  $-$28 49 21.91 $\pm$ 0.05  & $0.12 \pm 0.03$\tablefootmark{b}    & $0.18 \pm 0.03$    &  $0.17 \pm 0.03$  \\
        AR4   &  17 45 50.47 $\pm$ 0.05  &  $-$28 49 19.19 $\pm$ 0.05  & $0.34 \pm 0.03$\tablefootmark{b}    & $0.38 \pm 0.06$    &  $0.37 \pm 0.08$  \\
        AR5   &  17 45 50.57 $\pm$ 0.05  &  $-$28 49 17.23 $\pm$ 0.05  & $0.16 \pm 0.03$                     & $0.21 \pm 0.03$    &  $0.24 \pm 0.05$  \\
        AR6   &  17 45 49.76 $\pm$ 0.05  &  $-$28 49 25.62 $\pm$ 0.05  & $0.25 \pm 0.03$                     & $0.28 \pm 0.04$    &  $0.22 \pm 0.04$  \\
        AR7   &  17 45 50.83 $\pm$ 0.05  &  $-$28 49 26.05 $\pm$ 0.05  & $0.23 \pm 0.03$                     & $0.24 \pm 0.04$    &  $0.25 \pm 0.05$  \\
        AR8   &  17 45 50.45 $\pm$ 0.05  &  $-$28 49 31.56 $\pm$ 0.05  & $0.16 \pm 0.03$\tablefootmark{b}    & $0.20 \pm 0.03$    &  $0.22 \pm 0.05$  \\
        AR9   &  17 45 50.47 $\pm$ 0.05  &  $-$28 49 17.55 $\pm$ 0.06  & $0.15 \pm 0.04$                     & $0.22 \pm 0.03$    &  $0.06 \pm 0.01$  \\
        AR10  &  17 45 49.69 $\pm$ 0.05  &  $-$28 49 25.45 $\pm$ 0.05  & $0.06 \pm 0.03$                     & $0.14 \pm 0.02$    &  $0.16 \pm 0.03$  \\
        AR11  &  17 45 51.46 $\pm$ 0.05  &  $-$28 49 25.40 $\pm$ 0.05  &                                     & $0.08 \pm 0.01$    &  $0.06 \pm 0.01$  \\
        AR12  &  17 45 50.68 $\pm$ 0.05  &  $-$28 49 22.25 $\pm$ 0.06  &                                     & $0.07 \pm 0.01$    &  $0.07 \pm 0.02$  \\
        AR13  &  17 45 50.55 $\pm$ 0.05  &  $-$28 49 23.18 $\pm$ 0.05  &                                     & $0.16 \pm 0.03$    &  $0.23 \pm 0.05$  \\
        AR14  &  17 45 50.28 $\pm$ 0.05  &  $-$28 49 16.90 $\pm$ 0.05  &                                     & $0.05 \pm 0.01$    &  $0.09 \pm 0.02$  \\
        AR15  &  17 45 50.26 $\pm$ 0.05  &  $-$28 49 11.40 $\pm$ 0.05  &                                     & $0.10 \pm 0.02$    &  $0.27 \pm 0.06$  \\
        AR16  &  17 45 50.53 $\pm$ 0.05  &  $-$28 49 20.26 $\pm$ 0.07  &                                     &                    &  $0.04 \pm 0.01$  \\
        AR17  &  17 45 48.60 $\pm$ 0.05  &  $-$28 49 42.27 $\pm$ 0.05  &                                     & $0.22 \pm 0.03$    &  $0.23 \pm 0.05$  \\
        AR18  &  17 45 48.56 $\pm$ 0.05  &  $-$28 50 05.65 $\pm$ 0.05  &                                     & $0.17 \pm 0.03$    &  $0.18 \pm 0.04$  \\
        \noalign{\smallskip}
        \hline
        \end{tabular}
        \tablefoot{\smallskip \\
            \tablefoottext{a}{Units of right ascension are hours, minutes, and seconds, and units of declination are degrees, arcminutes, and arcseconds. Errors are in seconds and in arcseconds, respectively.} \\
        \tablefoottext{b}{Variable radio source according to \citet{lang:2005aa}.}
         }
        \label{table:point_sources}
\end{table*}

First, we noted that the 2018 X-band fluxes of the radio stars AR1, AR2, AR4, AR6, AR7, and AR8 (see fifth column in Table\,\ref{table:point_sources}) agreed well within their $1\,\sigma$ uncertainties with the $8.5$\,GHz flux densities (fourth column) reported by \citet{lang:2005aa}, assuming typical values for the spectral index, which indicated that the flux calibration of the 2018 image was the correct one. Subsequently we determined a secondary flux calibration factor for the 2016 X-band data by computing the weighted mean of the flux ratios of the previously mentioned sources between the two epochs. It was $1.8 \pm 0.1$. We cross checked this calibration factor in two ways: On the one hand, we followed the previous procedure, but excluded all sources labelled as variable by \citet{lang:2005aa}. With the remaining sources, AR2, AR6, and AR7, we obtained a mean flux ratio of $1.95 \pm 0.17$. On the other hand, we computed the weighted mean of the flux ratios of all stars detected in the 2016 and 2018 X-band images, except AR9, AR13, and AR15, which proved to be significant outliers. We found a calibration factor of $1.8 \pm 0.1$. After these consistency checks we applied a correction factor of $1.8 \pm 0.1$ to the 2016 X-band data. The re-calibrated 2016 flux densities are listed in the last column of Table\,\ref{table:point_sources}.

% End of Point Source Detection and Calibration subsection

\subsection{Flux densities and spectral indices}

% TABLE 4
\begin{sidewaystable*}[tph!]
  \centering
   \caption{JVLA flux densities of the compact sources from these observations.}
   \footnotesize
        \begin{tabular}{|l|l|l|c|c|c|c|r|l|}
            \hline
                \multicolumn{1}{| c |}{Source} & \multicolumn{1}{| c |}{Near-IR} & \multicolumn{2} {| c |} {Positions in X band (J2000.0)\tablefootmark{c}} & \multicolumn{3} {| c |} {Flux density (mJy)} & \multicolumn{1}{| c |}{$\alpha$} & \multicolumn{1}{| c |}{Spectral} \\
                  \multicolumn{1}{| c |}{Name\tablefootmark{a}} & \multicolumn{1}{| c |}{Counterpart\tablefootmark{b}}  & \multicolumn{1}{| c |}{R.A.} & Dec. & X-Band (2016) & X-Band (2018) & C-band (2018) & & \multicolumn{1}{| c |}{type} \\
                \hline
                
        AR1   & F6, Dong80\tablefootmark{d}  & 17 45 50.42 $\pm$ 0.05  & $-$28 49 21.95 $\pm$ 0.05  & $2.2   \pm 0.5$   & $2.1   \pm 0.3$   & $2.1   \pm 0.2$   & $-0.0 \pm 0.3$ & WN8-9h\tablefootmark{f}   \\
        AR2   & F8, Dong84                   & 17 45 50.39 $\pm$ 0.05  & $-$28 49 20.93 $\pm$ 0.05  & $0.23  \pm 0.05$  & $0.24  \pm 0.04$  & $0.14  \pm 0.02$  & $1.1  \pm 0.4$ & WN8-9h\tablefootmark{f}   \\ 
        AR3   & F1, Dong85\tablefootmark{d}  & 17 45 50.21 $\pm$ 0.05  & $-$28 49 21.91 $\pm$ 0.05  & $0.17  \pm 0.03$  & $0.18  \pm 0.03$  & $0.12  \pm 0.01$  & $0.8  \pm 0.4$ & WN8-9h\tablefootmark{f}   \\
        AR4   & F7, Dong83                   & 17 45 50.47 $\pm$ 0.05  & $-$28 49 19.19 $\pm$ 0.05  & $0.37  \pm 0.08$  & $0.38  \pm 0.06$  & $0.28  \pm 0.03$  & $0.6  \pm 0.4$ & WN8-9h\tablefootmark{f}   \\
        AR5   & F4, Dong81                   & 17 45 50.57 $\pm$ 0.05  & $-$28 49 17.23 $\pm$ 0.05  & $0.24  \pm 0.05$  & $0.21  \pm 0.03$  & $0.15  \pm 0.02$  & $0.7  \pm 0.4$ & WN7-8h\tablefootmark{f}   \\
        AR6   & F19                          & 17 45 49.76 $\pm$ 0.05  & $-$28 49 25.62 $\pm$ 0.05  & $0.22  \pm 0.04$  & $0.28  \pm 0.04$  & $0.43  \pm 0.04$  & $-0.8 \pm 0.3$ & O4-5 Ia\tablefootmark{f}  \\
        AR7   & F3, Dong82                   & 17 45 50.83 $\pm$ 0.05  & $-$28 49 26.05 $\pm$ 0.05  & $0.25  \pm 0.05$  & $0.24  \pm 0.04$  & $0.18  \pm 0.02$  & $0.6  \pm 0.4$ & WN8-9h\tablefootmark{f}   \\
        AR8   & F5, Dong17\tablefootmark{d}  & 17 45 50.45 $\pm$ 0.05  & $-$28 49 31.56 $\pm$ 0.05  & $0.22  \pm 0.05$  & $0.20  \pm 0.03$  & $0.16  \pm 0.02$  & $0.4  \pm 0.4$ & WN8-9h\tablefootmark{f}   \\
        AR9   & F18, Dong83                  & 17 45 50.47 $\pm$ 0.05  & $-$28 49 17.55 $\pm$ 0.05  & $0.06  \pm 0.01$  & $0.22  \pm 0.03$  & $0.28  \pm 0.03$  & $-0.5 \pm 0.3$ & O4-5 Ia$^{+}$\tablefootmark{f} \\
        AR10  & F2, Dong18                   & 17 45 49.69 $\pm$ 0.05  & $-$28 49 25.45 $\pm$ 0.05  & $0.16  \pm 0.03$  & $0.14  \pm 0.02$  & $0.12  \pm 0.01$  & $0.3  \pm 0.3$ & WN8-9h\tablefootmark{f}   \\
        AR11  & Dong79                       & 17 45 51.46 $\pm$ 0.05  & $-$28 49 25.40 $\pm$ 0.06  & $0.06  \pm 0.01$  & $0.08  \pm 0.01$  & $0.047 \pm 0.008$ & $1.0  \pm 0.4$ & WN7\tablefootmark{g}   \\       
        AR12  & F14, Dong88                  & 17 45 50.68 $\pm$ 0.05  & $-$28 49 22.25 $\pm$ 0.05  & $0.07  \pm 0.02$  & $0.07  \pm 0.01$  & < 0.025\tablefootmark{e}  & \multicolumn{1}{c |}{> 2.0}  & WN8-9h\tablefootmark{f}  \\
        AR13  & F26                          & 17 45 50.55 $\pm$ 0.05  & $-$28 49 23.18 $\pm$ 0.05  & $0.23  \pm 0.05$  & $0.16  \pm 0.03$  & < 0.020\tablefootmark{e}  & \multicolumn{1}{c |}{> 4.1}  & O4-5 Ia\tablefootmark{f} \\ 
        AR14  & F12, Dong87                  & 17 45 50.28 $\pm$ 0.05  & $-$28 49 16.90 $\pm$ 0.05  & $0.09  \pm 0.02$  & $0.05  \pm 0.01$  & $0.056 \pm 0.009$ & $-0.2 \pm 0.5$ & WN7-8h\tablefootmark{f}  \\
        AR15  & F9, Dong86                   & 17 45 50.26 $\pm$ 0.05  & $-$28 49 11.40 $\pm$ 0.05  & $0.27  \pm 0.06$  & $0.10  \pm 0.02 $ & $0.09  \pm 0.01$  & $0.2  \pm 0.4$ & WN8-9h\tablefootmark{f}  \\
        AR16  & F16                          & 17 45 50.53 $\pm$ 0.05  & $-$28 49 20.26 $\pm$ 0.07  & $0.04  \pm 0.01$  & < 0.037\tablefootmark{e} & < 0.023\tablefootmark{e}  & > 0.9  & WN8-9h\tablefootmark{f} \\
        AR17  & Dong19                       & 17 45 48.60 $\pm$ 0.05  & $-$28 49 42.27 $\pm$ 0.05  & $0.23  \pm 0.05$  & $0.22  \pm 0.03$  & $0.15  \pm 0.02$  & $0.7  \pm 0.4$ & WN8-9h\tablefootmark{g}   \\
        AR18  & Dong96                       & 17 45 48.56 $\pm$ 0.05  & $-$28 50 05.65 $\pm$ 0.05  & $0.18  \pm 0.04$  & $0.17  \pm 0.03$  & $0.11  \pm 0.01$  & $0.9  \pm 0.4$ & Ofpe/WN9\tablefootmark{g}   \\
 %        \noalign{\smallskip}
        \hline
        \end{tabular}
        \tablefoot{ \\ \\
    \tablefoottext{a}{Nomenclature for cluster members adopted by \citet[2005]{lang:2001aa}.}\\
    \tablefoottext{b}{Stellar identification as listed in \citet{clark:2018aa} and \citet{dong:2011aa}.} \\
    \tablefoottext{c}{Units of right ascension are hours, minutes, and seconds, and units of declination are degrees, arcminutes, and arcseconds. Errors are in seconds and in arcseconds, respectively.} \\
    \tablefoottext{d}{Variable radio source according to \citet{lang:2005aa}.} \\
    \tablefoottext{e}{Upper limits of the undetected sources fixed from the peak of the unresolved emission.} \\
    \tablefoottext{f}{Spectral classification by \citet{clark:2018aa}.} \\
    \tablefoottext{g}{Spectral classification as listed in Table 3 of \citet{dong:2011aa}.} \\
    }
    \label{table:results}
\end{sidewaystable*}

The X- and C-band flux densities for all radio stars measured in our 2016 and 2018 data are listed in Table\,\ref{table:results} (after secondary calibration of the 2016 X-band data, see above; the uncertainty of the calibration factor was included in the uncertainty of the reported flux densities).

The spectral index, $\alpha$, can be determined assuming that $S_{\nu}$ $\propto$ $\nu^{\alpha}$, where $\nu$ is the observing frequency and $S_{\nu}$ the measured flux density. We find  $\alpha\approx0.6$ for an isotropic and homogeneous wind with a constant terminal velocity, electron density, and chemical composition.

We inferred the spectral indices of the observed radio sources from their observed X- and C-band flux densities in 2018 using the expression

\begin{equation}
    \alpha = \frac {\log(S_{\nu_{\textrm{C}}}/S_{\nu_{\textrm{X}}})}{\log(\nu_{\textrm{C}}/\nu_{\textrm{X}})} \,,
\end{equation}
\vspace{0.2cm}

\noindent and the corresponding uncertainties were derived from standard error propagation as

\begin{equation}
    \sigma\alpha = \frac{1}{\log{(\nu_{\textrm{C}}/\nu_{\textrm{X}})}} \times \sqrt{\displaystyle\biggr(\frac{\sigma S_{\nu_{\textrm{X}}}}{S_{\nu_{\textrm{X}}}}\displaystyle\biggr)^{2} + \displaystyle\biggr(\frac{\sigma S_{\nu_{\textrm{C}}}}{S_{\nu_{\textrm{C}}}}\displaystyle\biggr)^{2}},
\end{equation}
\vspace{0.2cm}

\noindent where terms preceded by $\sigma$ refer to the standard deviation.
The computed spectral indices and their uncertainties are listed in the second to last column of Table\,\ref{table:results}.

% End of Flux Densities and Spectral Indices subsection

\subsection{Mass-loss rates}

Most methods to estimate the mass-loss rate (in ultraviolet, optical, or infrared regimes) suffer from the uncertainty that the derived rates depend on non-observable parameters. Nevertheless, observations of radio continuum radiation yield mass-loss rates based only on observable quantities: the radio flux density, the terminal velocity of the stellar wind, and the distance to the star. According to \citet{panagia:1975aa} and \citet{wright:1975aa}, the mass-loss rates in solar masses per year can be written as

\begin{eqnarray}
    \lefteqn{\displaystyle\biggl[ \frac{\dot{M}}{\textrm{M}_{\odot}\,\textrm{yr$^{-1}$}}\displaystyle\biggr] = 5.34 \times 10^{-4} \displaystyle\biggl[ \frac{S_{\nu}}{\textrm{mJy}}\displaystyle\biggr]^{3/4} \displaystyle\biggl[ \frac{v_{\infty}}{\textrm{Km\,s$^{-1}$}}\displaystyle\biggr] \displaystyle\biggl[ \frac{d}{\textrm{kpc}}\displaystyle\biggr]^{3/2}}
    \nonumber\\
    & & {} \hspace{1.7cm} \displaystyle\biggl[ \frac{\nu}{\textrm{Hz}}\displaystyle\biggr]^{-1/2} \displaystyle\biggl[ \frac{{\mu}^{2}}{Z \gamma g_{\nu}}\displaystyle\biggr]^{1/2} \,,
\end{eqnarray}
\vspace{0.2cm}

\noindent where $S_{\nu}$ is the flux density in milli-Janskys, $v_{\infty}$ is the terminal velocity of the stellar wind in kilometers per second, $\nu$ is the observed frequency in Hertz, and $d$ is the distance to the observer in kiloparsecs ($\sim$ 8\,kpc in our case). The parameters $\mu$, $Z,$ and $\gamma$ are the mean molecular weight, the mean ionic charge, and the mean number of electrons per ion. We adopted the values of $v_{\infty}$ from Table 2 by \citet{martins:2008aa} and from their He/H values we have calculated $\mu$ using the approximation

\begin{equation}
    \mu = \frac{\sum_{j}n_{j} A_{j}}{\sum_{j}n_{j}}  \,,
\end{equation}
\vspace{0.2cm}

\noindent where $n_{j}$ is the number density of atoms of type j, and $A_{j}$ is the mass number. We only considered H and He, neglecting any other metals in the winds, and so the expression can be simplified as

\begin{equation}
    \mu= \frac{1\, + 4 \cdot \textrm{He/H}}{1\, + \textrm{He/H}}  \,,
\end{equation}
%\vspace{0.2cm}

\noindent where He/H is the ratio of H to He given in Table 2 by \citet{martins:2008aa}. We note that AR6, AR11, AR17, and AR18 are not included in Table 2 by \citet{martins:2008aa}, and so we adopted values of stars of similar type. $Z$ and $\gamma$ are parameters that depend on the ionisation conditions in the wind, which leads to $Z = \gamma = 1$ in a radio-emitting region of WR stars \citep[see][]{leitherer:1997aa}.

The free-free Gaunt factor, $g_{\nu}$, can be obtained by means of the expression

\begin{equation}
    g_{\nu} = 9.77 \cdot \displaystyle\biggl(1 + 0.13 \cdot \textrm{log}\,\frac{T_\mathrm e^{3/2}}{Z \nu}\displaystyle\biggr) \,,
\end{equation}
%\vspace{0.2cm}

\noindent using the approximation by \citet{leitherer:1991aa}, where $T_{\mathrm e}$ is the electron temperature of the wind in kelvins.
We assumed $T_{\mathrm e} = 10^{4}$\,K (deviations from this temperature had only minor effects on $g_{\nu}$).

We now know the distance of Sgr\,A* to $< 0.5\%$. As concerns the line-of-sight distance of Arches with respect to Sgr\,A*, we can assume that it is located within the circum molecular zone, that is $r <= 200$\,pc. Therefore, assuming a 200\,pc uncertainty on the distance of Sgr\,A* and a 200\,pc uncertainty of Arches relative to Sgr\,A*, we get a relative uncertainty of $0.4\%$. This uncertainty is systematic in the sense that it affects all sources in Arches in the same way. The distribution of sources within Arches is irrelevant because we are dealing with at most a few parsecs difference. Hence, the uncertainty of $d$ can be neglected. For the remaining parameters, we estimated the uncertainties using the same criteria as in \citet{leitherer:1997aa}. A $10\%$ for $v_{\infty}$, and also a $10\%$ for the free-free Gaunt factor. The assumed uncertainties for $Z$, $\gamma,$ and $\mu$ are $\pm$0.08\,dex. Therefore, according to the standard error propagation and defining $k = 5.34 \times 10^{-4}$, the uncertainties were obtained from

\begin{equation}
    \begin{split}
        \alpha(\dot{M}) & = \displaystyle\biggl[\, \frac{9}{16}\,\displaystyle\biggr( \frac{\dot{M}}{k\cdot\,S_{\nu}}\displaystyle\biggr)^{2}\,(\alpha\,S_{\nu})^{2} \,+ \displaystyle\biggl( \frac{\dot{M}}{k\cdot\,v_{\infty}}\displaystyle\biggr)^{2} (\alpha\,v_{\infty})^{2}\, + \\
        & + \frac{9}{4}\,\displaystyle\biggr( \frac{\dot{M}}{k\cdot\,d}\displaystyle\biggr)^{2}\, (\alpha\,d)^{2} + 
        \frac{1}{4}\,\displaystyle\biggr( \frac{\dot{M}}{k\cdot\,\nu}\displaystyle\biggr)^{2}\, (\alpha\,\nu)^{2} \,+ \\
        & + \displaystyle\biggl( \frac{\dot{M}}{k\cdot\,\mu}\displaystyle\biggr)^{2}\,(\alpha\,\mu)^{2} + \frac{1}{4}\,\displaystyle\biggr( \frac{\dot{M}}{k\cdot\,Z}\displaystyle\biggl)^{2}\,(\alpha\,Z)^{2} + \\
        & + \frac{1}{4}\,\displaystyle\biggr( \frac{\dot{M}}{k\cdot\,\gamma}\displaystyle\biggl)^{2}\,(\alpha\,\gamma)^{2} +
        \frac{1}{4}\,\displaystyle\biggr( \frac{\dot{M}}{k\cdot\,g_{\nu}}\displaystyle\biggl)^{2}\,(\alpha\,g_{\nu})^{2}\, \displaystyle\biggr]^{1/2} \,,
    \end{split}
\end{equation}

\vspace{0.2cm}

\noindent where terms preceded by $\sigma$ also refer to the standard deviation, and we obtained a typical error of the order of 10$^{-6}$ on log\,$\dot{M}$.

% TABLE 5
\begin{table*}[th!]
        \caption{Radio mass-loss rates in the C-band (2018) and in the two X-band epochs (2016/2018), where the typical logarithmic error is $\pm$\,0.2\,dex on log\,$\dot{M}$, and comparison to estimations made by Lang et al. from their 8.5\,GHz observations \citep{lang:2001aa} and their 22.5\,GHz observations \citep{lang:2005aa}, both re-scaled for taking into account their assumed $v_{\infty}$ = 1000\,Km\,s$^{-1}$, and comparison to estimations made by \citet{martins:2008aa}, also re-scaled considering their clumping factor value of 0.1. All mass-loss rates, $\dot{M}$, are in units of M$_{\odot}$\,yr$^{-1}$.}
        \small
        \setlength{\tabcolsep}{6.9pt}  % Default value: 6pt
        \begin{tabular}{lccccccccc}
         \hline
         \noalign{\smallskip}
         Source & He/H\tablefootmark{a} & $v_{\infty}$\tablefootmark{a} & $\mu$\tablefootmark{b} & $\dot{M}_{2018}$  & $\dot{M}_{2016}$   & $\dot{M}_{2018}$   & $\dot{M}_{\textrm{Lang\,2001}}$  & $\dot{M}_{\textrm{Lang\,2005}}$                      & $\dot{M}_{\textrm{Martins\,2008}}$  \\
        \noalign{\smallskip}
                & (\#)                  & (Km/s)                        & (\#)                   & 6.0\,GHz          & 10.0\,GHz          & 10.0\,GHz          & 8.5\,GHz                & 22.5\,GHz\tablefootmark{c}          & IR K-Band\tablefootmark{d}  \\
     \noalign{\smallskip}
         \hline \hline
     \noalign{\smallskip}
        AR1   & 0.2  & 1400  & 1.5 & $2.1 \times 10^{-4}$ & $2.1 \times 10^{-4}$ & $2.0 \times 10^{-4}$  & $2.4 \times 10^{-4}$ & $3.1 \times 10^{-4}$ & $7.6 \times 10^{-5}$ \\
        AR2   & 1.0  & 1000  & 2.5 & $3.2 \times 10^{-5}$ & $4.6 \times 10^{-5}$ & $4.9 \times 10^{-5}$  & $3.9 \times 10^{-5}$ & $4.6 \times 10^{-5}$ & $1.0 \times 10^{-4}$ \\ 
        AR3   & 0.1  & 1400  & 1.3 & $2.1 \times 10^{-5}$ & $2.7 \times 10^{-5}$ & $2.8 \times 10^{-5}$  & $4.5 \times 10^{-5}$ & $2.7 \times 10^{-5}$ & $6.3 \times 10^{-5}$ \\
        AR4   & 0.3  & 1300  & 1.7 & $4.7 \times 10^{-5}$ & $5.9 \times 10^{-5}$ & $5.9 \times 10^{-5}$  & $5.1 \times 10^{-5}$ & $6.2 \times 10^{-5}$ & $7.9 \times 10^{-5}$ \\
        AR5   & 0.4  & 1400  & 1.9 & $3.6 \times 10^{-5}$ & $5.0 \times 10^{-5}$ & $4.6 \times 10^{-5}$  & $4.2 \times 10^{-5}$ & $2.7 \times 10^{-5}$ & $1.4 \times 10^{-4}$ \\
        AR6   & 0.1  & 2400  & 1.3 & $9.2 \times 10^{-5}$ & $5.5 \times 10^{-5}$ & $6.8 \times 10^{-5}$  & $1.1 \times 10^{-4}$ & $7.2 \times 10^{-5}$ & \\
        AR7   & 0.6  &  800  & 2.1 & $2.6 \times 10^{-5}$ & $3.3 \times 10^{-5}$ & $3.2 \times 10^{-5}$  & $3.4 \times 10^{-5}$ & $3.4 \times 10^{-5}$ & $7.9 \times 10^{-5}$ \\
        AR8   & 0.8  &  900  & 2.3 & $3.0 \times 10^{-5}$ & $3.7 \times 10^{-5}$ & $3.4 \times 10^{-5}$  & $3.2 \times 10^{-5}$ & $3.3 \times 10^{-5}$ & $7.3 \times 10^{-5}$ \\
        AR9   & 0.1  & 2150  & 1.3 & $5.9 \times 10^{-5}$ & $2.0 \times 10^{-5}$ & $5.1 \times 10^{-5}$  &                      & $4.1 \times 10^{-5}$ & $1.4 \times 10^{-5}$ \\
        AR10  & 0.35 & 1400  & 1.8 & $2.9 \times 10^{-5}$ & $3.5 \times 10^{-5}$ & $3.2 \times 10^{-5}$  &                      & $2.7 \times 10^{-5}$ & $6.0 \times 10^{-5}$ \\
        AR11  & 0.4  & 1400  & 1.9 & $1.4 \times 10^{-5}$ & $1.9 \times 10^{-5}$ & $2.1 \times 10^{-5}$  &   &   & \\
        AR12  & 0.1  & 1400  & 1.3 &                      & $1.4 \times 10^{-5}$ & $1.3 \times 10^{-5}$  &   &   & $3.2 \times 10^{-5}$ \\
        AR13  & 0.1  & 2600  & 1.3 &                      & $6.3 \times 10^{-5}$ & $4.8 \times 10^{-5}$  &   &   & $6.0 \times 10^{-6}$ \\
        AR14  & 0.2  & 1500  & 1.5 & $1.4 \times 10^{-5}$ & $2.1 \times 10^{-5}$ & $1.4 \times 10^{-5}$  &   &   & $5.7 \times 10^{-5}$ \\
        AR15  & 0.1  & 1800  & 1.3 & $2.1 \times 10^{-5}$ & $4.9 \times 10^{-5}$ & $2.4 \times 10^{-5}$  &   &   & $5.4 \times 10^{-5}$ \\
        AR16  & 0.1  & 1400  & 1.3 &                      & $9.0 \times 10^{-6}$ &                       &   &   & $2.5 \times 10^{-5}$ \\
        AR17  & 0.4  & 1300  & 1.9 & $3.3 \times 10^{-5}$ & $4.6 \times 10^{-5}$ & $4.4 \times 10^{-5}$  &   &   & \\
        AR18  & 0.4  & 1300  & 1.9 & $2.6 \times 10^{-5}$ & $3.9 \times 10^{-5}$ & $3.6 \times 10^{-5}$  &   &   & \\
        \noalign{\smallskip}
        \hline
        \end{tabular}
        \tablefoot{\smallskip \\
        \tablefoottext{a}{Wind parameters from Table 2 by \citet{martins:2008aa}.} \\
        \tablefoottext{b}{Mean molecular weight determined using the equation (5) from this paper.} \\
        \tablefoottext{c}{Derived from the 22.5\,GHz flux density for AR1, AR2, AR4, AR7 and AR8 sources, and from the 8.5\,GHz flux density for the others, as described in Table 5 by \citet{lang:2005aa}.}\\
        \tablefoottext{d}{Infrared K-band obtained with the integral field spectrograph SINFONI on the VLT by \citet{martins:2008aa}. The typical error provided is $\pm$\,0.2\,dex on log\,$\dot{M}$.}
        }
        \label{table:mass-loss}
\end{table*}

With these premises, we derived mass-loss rates corresponding to the observed flux densities at 10\,GHz assuming that the observed radio emission is due to free-free emission from ionised extended envelopes with a steady and completely ionised wind, with a volume filling factor of $f = 1$, and an electron density profile of $n_{\textrm{e}}$ $\propto$ $r^{-2}$. In the case of non-thermal contributions, our values, shown in Table\,\ref{table:mass-loss}, represent upper limits to the true mass-loss rates.

Table\,\ref{table:mass-loss} also shows the estimations done by Lang et al. from their 8.5\,GHz observations \citep{lang:2001aa} and their 22.5\,GHz observations \citep{lang:2005aa}, and the estimations done by \citet{martins:2008aa}. Lang et al. assumed a terminal velocity of the wind of 1000\,Km\,s$^{-1}$ for all sources, and Martins et al. adopted a volume filling factor of 0.1. Therefore, in order to compare our data with their data, we re-scaled Lang's values multiplying them by $v_{\infty}$/1000, where $v_{\infty}$ are the values adopted in this paper, and we multiplied Martins' mass-loss rates by $1/\sqrt{0.1}$ \citep[see][]{abbott:1981aa} considering the clumpling factor.

\section{Discussion and conclusions}

\subsection{Properties of the sources}

All detected sources are young, massive stars that have evolved off the main sequence, with the great majority identified as Wolf-Rayet stars of type WNh (see Table\,\ref{table:results}), a classification that \citet{figer:1999ab, figer:2002aa} and recently \citet{clark:2018aa} have assigned to most of the Arches cluster members.

Even though the uncertainties on the inferred spectral indices are relatively high, we can say that the stars AR2, AR3, AR4, AR5, AR7, AR11, AR15, AR17, and AR18 have spectral indices consistent with thermal optically thick emission from ionised stellar winds. Small differences with respect to the canonical value could come from changes in the ionisation fraction along the emitting regions and/or wind structures such as clumps and/or shocks resulting from internal instabilities. The stars AR1, AR6, and AR9 have flat or inverted spectral indices, which may indicate that they are CWBs. AR10 has an inverted spectrum and is a confirmed CWB \citep[see][]{lohr:2018aa}. Ambiguous cases are the stars AR8 and AR14. No measurements of $\alpha$ are available for the stars AR12, AR13, and AR16 because they are only detected in a single band.

AR1 and AR4 are associated with X-ray sources, where the emission is suggested to arise in shocks in their winds \citep{lang:2005aa,law:2004aa}. Such shocks may arise in CWBs. The flat radio spectrum of AR1 (this work and \citealt{lang:2005aa}) supports this interpretation, but the close-to-thermal spectrum of AR4 (this work and \citealt{lang:2005aa}) does not provide any evidence for a CWB. Possibly, AR4 is a binary that is too tight to be identified as such by radio observations, as explained in the introduction. We find no significant variability of AR1 and AR4, as could be expected for highly eccentric CWBs observed at significantly different phases of their orbits. The nature of the X-ray emission in both sources requires further investigation.

As we have two epochs with X-band measurements, we can probe variability. AR16 is the weakest stellar source detected by our observations. It lies near the crowded centre of the cluster and close to the brightest source, AR1. Its non-detection in 2018 is therefore probably rather due to the lower quality of the 2018 data than to variability. In order to have been detectable in the 2018 data its flux would have had to increase by a factor of three in brightness (0.012 vs. 0.04\,mJy). From the 17 sources detected in two epochs, only AR9 and AR15 display unambiguous variability with $>5\sigma$ significance. In particular, source AR9 is labelled as a new detection in \citet{lang:2005aa}, as compared to \citet{lang:2001aa}. Therefore, its variability can be considered to be firmly established. We note that AR9 has a negative spectral index, and so binarity and its associated orbital motion are possible causes for its radio variability. The X-band fluxes of the other sources agree within $1\,\sigma$ (12 sources) or $2\,\sigma$ (3 sources). \citet{lang:2005aa} mark the four sources AR1, AR3, AR4, and AR8 as potentially variable. When we compare our measurements with theirs (propagating the 10\,GHz flux densities to $8.5$\,GHz with the help of the measured spectral indices), we do not find any signs for variability within the uncertainties on our measurements. In conclusion, if we omit the ambiguous case of AR16, we find that only 2 of the 17 radio stars, or less than 15\%, display significant variability. The radio emission therefore appears to be stable on timescales of a few to ten years.

\subsection{Number of detected sources}

% TABLE 6
\begin{table*}[ht!]
\caption{Expected number of stars detected at 10\,GHz within cylindrical beam of $R=0.5$\,pc, assuming a mass of $1\times10^{4}$\,M$_{\odot}$, a volume filling factor of 1, different ages and two different slopes $a$ of the initial mass function. The numbers in brackets correspond to the assumption of a volume filling factor of $0.1$ for the stellar wind.}
\label{tab:n_expected}
\small
\setlength{\tabcolsep}{12pt}  % Default value: 6pt
 
\begin{tabular}{l | llll}
  Age (Myr)           & 2 & 3   &   4 & 5 \\
\noalign{\smallskip}
\hline
\hline
\noalign{\smallskip}
 PARSEC & & & & \\
\noalign{\smallskip}
\hline
\noalign{\smallskip}
  $a=2.35$   & $1.4\pm1.2$ ($7.7\pm2.6$)  & $4.7\pm2.0$ ($6.4\pm2.3$)    &  $5.6\pm2.7$ ($5.9\pm2.7$)   & $3.4\pm1.7$ ($5.6\pm2.4$)   \\
  $a=1.8$    & $5.7\pm2.1$ ($26.2\pm4.8$) & $17.6\pm4.0$ ($22.6\pm4.6$)  &  $15.3\pm3.7$ ($16.2\pm3.9$) & $7.9\pm2.8$ ($12.8\pm3.4$)  \\
\noalign{\smallskip}
\hline
\noalign{\smallskip}
 MIST & & & & \\
\noalign{\smallskip}
\hline
\noalign{\smallskip}
  $a=2.35$   & $2.5\pm1.7$ ($8.1\pm3.2$)  & $5.8\pm2.4$ ($9.6\pm3.5$)    &  $3.3\pm2.0$ ($3.5\pm2.0$)   & $3.1\pm1.8$ ($3.1\pm1.8$)  \\
  $a=1.8$    & $9.6\pm3.1$ ($27.1\pm4.5$) & $20.6\pm4.2$ ($31.4\pm5.3$)  &  $10.1\pm3.1$ ($10.5\pm3.1$) & $7.8\pm2.9$ ($7.9\pm2.8$)  \\
\noalign{\smallskip}
\hline
\end{tabular}
\end{table*}

All but two of the 18 radio stars detected by our study lie within a projected radius of about $R=12.5"$ ---or $0.5$\,pc at the distance of the GC--- of the centre of the Arches cluster. It is possible that the sources far from the centre, AR17 and AR18, also originated in the cluster, a hypothesis that could be tested with proper motion measurements.

Within $R=0.5$\,pc we detect six more sources than the previous study by \citet{lang:2005aa}. The faintest source reported by \citet{lang:2005aa} is AR10 as a 2\,$\sigma$ detection with $0.06$\,mJy at $8.5$\,GHz. With the spectral indices estimated here (or assuming a flat spectral index as a conservative estimate where no measurement is available) we can estimate the $8.5$\,GHz flux density of the new sources found by our work. The sources AR11 ($0.06$\,mJy), AR12 ($\leq0.07$\,mJy), AR14 ($\leq0.09$\,mJy), and AR16 ($\leq0.04$\,mJy) therefore probably escaped from previous detection because of the lower sensitivity of the observations. AR15 is clearly identified as a variable source by our observations (see previous section). Also, AR11 and AR15 lie outside of the FoV shown in Fig.\,6 of \citet{lang:2005aa}. The relatively bright source AR13 should have been detected by \citet{lang:2005aa}. Its non-detection may be due to variability, possibly combined with its closeness to the brightest source, AR1, which may lead to it being confused with side lobes from this source.

Does the number of detected radio sources correspond to our expectations? First of all, we detect all spectroscopically classified WR stars in the cluster \citep[see][]{clark:2018aa}. Our sample of these sources of strong ionised winds is therefore complete. The faintest source detected by the highest quality observations of our work is AR16 with a 10\,GHz flux density of $0.04\pm0.01$\,mJy. Its mass-loss rate is approximately $9.0\times10^{-6}$\,M$_{\odot}$\,yr$^{-1}$ or about $2.8\times10^{-6}$\,M$_{\odot}$\,yr$^{-1}$ if we assume a volume filling factor of $0.1$ for the stellar wind and a wind velocity of $v_{\infty}=1400$\,km\,s$^{-1}$, which are typical values for our sources.

We can estimate the number of stars with a wind mass-loss rate at or above this value by combining estimates of the age, mass, and initial mass function (IMF) of the Arches cluster with isochrones of stellar evolutionary codes. For the IMF we assume a one-segment power-law IMF, using either the standard Salpeter exponent of $2.35$ or the observationally motivated top-heavy value of $1.8$ \citep{hosek:2019aa}. The latter value is also in good agreement with the observational value of the present day mass function of the cluster within about $R=0.5$\,pc of its centre \citep{stolte:2005aa}. All stars that are clearly associated with the cluster, namely AR1 to AR16, are located within a projected radius of $R\approx0.5$\,pc of the cluster centre. Following \citet{clarkson:2012aa} and \citet{hosek:2019aa}, we can estimate the cluster mass within this area to $1\times10^{4}$\,M$_{\odot}$ with an approximate uncertainty of 30\%.  Using PARSEC \citep[release v$1.2$S $+$ COLIBRI S$\_35$][]{Bressan:2012aa,Chen:2014aa,Chen:2015aa,Tang:2014aa,Marigo:2017aa,Pastorelli:2019aa} and MIST \citep{Dotter:2016aa,Choi:2016aa,Paxton:2011aa,Paxton:2013aa,Paxton:2015aa} theoretical isochrones for solar metallicity we can estimate the number of stars with a mass-loss rate of $\gtrsim$\,9.0$\times10^{-6}$\,M$_{\odot}$\,yr$^{-1}$ ($\gtrsim 2.8\times10^{-6}$\,M$_{\odot}$\,yr$^{-1}$ for a $0.1$ volume filling factor) in this area.

Table\,\ref{tab:n_expected} lists this number for the two IMF slopes assumed here, for five different cluster ages, for the two theoretical models, and for assuming volume filling factors of 1 and 0.1. In each case, the numbers are the means and standard deviations from 100 runs of a Monte Carlo simulation. The numbers for MIST and PARSEC agree within their estimated uncertainties. The assumed volume filling factor has little impact at ages of 4 and 5\,Myr, but can increase the numbers of detected stars considerably at lower ages.

Comparing the numbers in Table \ref{tab:n_expected} with the number of detected radio stars in our observations, either 16 or 18, the latter if we assume that AR17 and AR18 may be high-mass stars that escaped from the cluster, we can see that an age as old as 5\,Myr can be safely ruled out. The numbers clearly also require a top-heavy IMF; they indicate an age of the cluster in the range 2-4\,Myr. This agrees well with the observational values of the age range: from  $3.7\pm0.2$\,Myr \citep{hosek:2019aa} and $3.7\pm0.7$\,Myr  \citep{schneider:2014aa} on the high end, to lower estimates of around $2.0-3.3$\,Myr \citep[e.g.][]{clark:2018aa,lohr:2018aa}. 

Our age estimation is relatively crude because it relies on various assumptions ---among others the validity of theoretical codes for the post-main sequence evolution of massive stars--- and ignores complications such as the influence of stellar multiplicity on stellar evolution.  Nevertheless, our estimate shows that the number of radio stars can be used as a tool to constrain the age or mass of a cluster and also its mass function. We also show that the theoretical models appear to satisfactorily represent mass-loss rates of massive stars.
 
We  show that with the extraordinary sensitivity of the JVLA we can detect massive stars with mass-loss rates up to $1.3\times 10^{-5}$\,M$_{\odot}$\,yr$^{-1}$ in Arches. With the advent of the Square Kilometre Array at mid-frequency (SKA-MID), which will cover the frequency range 350\,MHz to 14\,GHz, especially at the highest frequencies (Band 5) where the highest contribution from the optically thick stellar wind is expected, it will be possible to measure mass-loss rates of $\sim\,10^{-7}$\,M$_{\odot}$\,yr$^{-1}$ at the distance of the GC in tens of minutes integration time. This would favour studies of massive stars and their associated winds at all stages of evolution, including pre-main sequence stars, luminous blue variables, and Wolf Rayet stars, and at the same age inside the three main massive stellar clusters at the GC (the Central Cluster, Arches, and Quintuplet).

\vspace{0.5cm}

\begin{acknowledgements}
      Karl G. Jansky Very Large Array (JVLA) of the National Radio Astronomy Observatory (NRAO) is a facility of the National Science Foundation (NSF) operated under cooperative agreement by Associated Universities, Inc.\\
      A.~T. G.-C., R. S., A. A. and B. S. acknowledge financial support from the State Agency for Research of the Spanish MCIU through the ``Center of Excellence Severo Ochoa'' award for the Instituto de Astrof\'isica de Andaluc\'ia (SEV-2017-0709). \\
      A.~T. G.-C., R. S., and B. S. acknowledge financial support from national project PGC2018-095049-B-C21 (MCIU/AEI/FEDER, UE).\\
      A. A. acknowledges support from national project PGC2018-098915-B-C21 (MCIU/AEI/FEDER, UE). \\
      F. N. acknowledges financial support through Spanish grants ESP2017-86582-C4-1-R and PID2019-105552RB-C41 (MINECO/MCIU/AEI/FEDER) and from the Spanish State Research Agency (AEI) through the Unidad de Excelencia ``Mar\'ia de Maeztu''-Centro de Astrobiolog\'ia (CSIC-INTA) project No. MDM-2017-0737. \\
      F. N.-L. gratefully acknowledges funding by the Deutsche Forschungsgemeinschaft (DFG, German Research Foundation) -- Project-ID 138713538 -- SFB 881 (``The Milky Way System'', subproject B8).\\
      The research leading to these results has received funding from the European Research Council under the European Union's Seventh Framework Programme (FP7/2007-2013) / ERC grant agreement n$^{\circ}$ [614922].
\end{acknowledgements}

\bibliographystyle{aa} % style aa.bst
\bibliography{Arches} % your references Yourfile.bib

\end{document}